%% file: Infinitary_Logic_Universal_CPS_TARK__Colored_Hyperlinks_.tex
\documentclass[submission,copyright,creativecommons]{eptcs}

\providecommand{\event}{TARK2017}

\usepackage{mystyle_article}

\title{The Topology-Free Construction of the Universal Type Structure for Conditional Probability Systems}
\author{Pierfrancesco Guarino
\institute{School of Business and Economics\\
Maastricht University (AE1)\\
Maastricht, The Netherlands}
\email{p.guarino@maastrichtuniversity.nl}%
}
\def\titlerunning{The Topology-Free Universal Type Structure for CPSs}
\def\authorrunning{Pf. Guarino}
\begin{document}

\maketitle

\input{./Background/abstract.tex}


\input{./Text/Introduction/introduction.tex}

\input{./Text/Introduction/terminology_new.tex}

\input{./Text/Introduction/synopsis.tex}


\input{./Text/preliminaries.tex}


\input{./Text/Type_Structures/intro.tex}

\input{./Text/Type_Structures/standard.tex}

\input{./Text/Type_Structures/logic.tex}

\input{./Text/Type_Structures/large.tex}


\input{./Text/Terminality/terminality.tex}

\input{./Text/Terminality/infinite_hierarchies.tex}


\input{./Text/non-redundancy.tex}


\input{./Text/Belief-Completeness/intro.tex}

\input{./Text/Belief-Completeness/infinitary_logic/intro.tex}

\input{./Text/Belief-Completeness/infinitary_logic/syntax.tex}

\input{./Text/Belief-Completeness/infinitary_logic/semantics.tex}

\input{./Text/Belief-Completeness/strong_soundness_completeness/intro.tex}

\input{./Text/Belief-Completeness/strong_soundness_completeness/strong_soundness.tex}

\input{./Text/Belief-Completeness/strong_soundness_completeness/strong_completeness.tex}

\input{./Text/Belief-Completeness/belief-completeness.tex}

%
\input{./Background/bibliography.tex}

\end{document}

%% file: Background/abstract.tex
\begin{abstract}
We construct the universal type structure for conditional probability systems without any topological assumption, namely a type structure that is terminal, belief-complete,  and non-redundant. In particular, in order to obtain the belief-completeness in a constructive way, we extend the work of Meier [An Infinitary Probability Logic for Type Spaces. Israel Journal of Mathematics, 192, 1--58] by proving strong soundness and strong completeness of an infinitary conditional probability logic with truthful and non-epistemic conditioning events.
\end{abstract}

\vspace{-0.5cm}

%% file: Text/Introduction/introduction.tex
\section{Introduction}
\label{sec:introduction}

Games with incomplete information are defined as games where there is lack of ``common knowledge'' concerning some aspects of the interaction under scrutiny. Historically the analysis of such games proved to be problematic due to the infinite regress of beliefs that they induce. That is, starting from a parameter space $X$, a player forms beliefs concerning $X$ ($1^{\text{st}}$-order beliefs), beliefs concerning $X$ \emph{and} what the other players believe about $X$ ($2^{\text{nd}}$-order beliefs), beliefs concerning $X$ \emph{and} what the other players believe about $X$ and what she believes about $X$ ($3^{\text{rd}}$-order beliefs), and so on, where the final object comprised of all these belief orders is called a \emph{belief hierarchy}. The problem of obtaining, in line with the Bayesian paradigm, a single probability measure that describes the players' uncertainty about all those layers simultaneously hampered the possibility of implementing equilibrium analysis on this class of games.

The problem was solved in \cite{Harsanyi_1967} with the introduction of type structures.  A type of a player is an object that \emph{implicitly} contains all the information needed in order to retrieve the belief hierarchy, that is, formally a type of a player induces a probability measure over the parameter space and the types of the opponents. Hence, Harsanyi's intuition was that, as soon as the sets of types of all players are assumed to be ``common knowledge'', that is, they enter in the formal representation of the game, it becomes possible for every player to have a single probability measure over the aspects of the game not ``common knowledge'', thus being in position to perform equilibrium analysis.

Still, this solution, even if particularly handy in dealing with specific applications, left open two questions:
\begin{enumerate}
\item Is possible to actually construct a type with \emph{bare hands} from the infinite regress that we face in games with incomplete information?
\item Is possible to construct a type structure that, given a certain parameter space, contains \emph{all} possible beliefs that players can have about the parameter space and the beliefs of the other players?
\end{enumerate}

The answer to both questions came in the affirmative with \cite{Mertens_Zamir_1985},\footnote{\label{foot:Arm}Previous papers on the topic, which went unnoticed at the time, were \cite{Armbruster_Boge_1979} and \cite{Boge_Eisele_1979}.} that, starting from a compact Hausdorff parameter space,  \emph{explicitly} constructed a space that answers question (2). Moreover, their construction was based on the intuition that types are an \emph{implicit} way to represent infinite hierarchies of beliefs that have one fundamental property, namely of being \emph{coherent}, that is, higher order beliefs agree with lower order beliefs. This property allowed \cite{Mertens_Zamir_1985}  to construct types \emph{explicitly}, hence answering in the affirmative to question (1).

A rich literature arose on the construction of such large type structures from parameter spaces with different topological assumptions answering the previous two questions. However, the most general case, namely the measure-theoretic case without any topological assumptions showed itself to be elusive. The solution was provided by the path-breaking \cite{Heifetz_Samet_1998}, that introduced two alternative constructions -- both different from the standard arguments relying on coherency -- of a topology-free structure that answers question (2). Indeed, with \cite{Heifetz_Samet_1999} it also became clear that in this case it is not possible to identify the set of all types with the set of all coherent belief hierarchies thus explicitly answering question (1). Hence, the main idea of \cite{Heifetz_Samet_1998} was to establish the existence, given a certain parameter space, of a type structure $\mscr{T}^*$ -- which is what we call in this paper the \emph{terminal} type structure, while they called it \emph{universal}\footnote{See \Sref{subsec:terminology} for an explanation of why we choose to call such space ``terminal''.} -- with the property that any other type structure $\mscr{T}$ can be uniquely embedded into it in a formally appropriate sense. Thus, in following this path, the authors moved back to the the \emph{implicit} approach \emph{\'a la} Harsanyi.

Starting from a Polish parameter space, \cite{Battigalli_Siniscalchi_1999} obtained an \emph{explicit} construction of what we call a universal type structure for the case of conditional probability systems.\footnote{Conditional probability systems have been introduced in the game-theoretic literature by \cite{Myerson_1986} building on the notion of \emph{conditional probability space} of \cite{Renyi_1955} (see \cite{Halpern_2010} for an analysis of this and related notions).} The authors also made a conjecture concerning the possibility to perform a topology-free construction \emph{\'a la} \cite{Heifetz_Samet_1998} for conditional probability systems. The present paper answers in the affirmative to this longstanding conjecture, thus proving the existence of the terminal and non-redundant type structure with conditioning events for the purely measure-theoretic case. Moreover, we \emph{explicitly}\footnote{Observe that it is the fact that this construction is indeed explicit (i.e., performed via infinitary probability logic) that makes it more informative than constructions performed via coalgebraic methods, since both constructions ensure that the type structure obtained is both terminal and belief-complete ((see \Sref{subsec:large_type_structures} for an explanation of these notions). Indeed, as noticed by \cite{Heinsalu_2013}, proving the terminality of a type structure via coalgebraic methods ensures \emph{a fortiori} also the belief-completeness of this type structure thanks to a standard result of category theory  from \cite{Lambek_1968} known as Lambek's Lemma.} construct a topology-free type structure, which we prove to be isomorphic to the terminal one, which is belief-complete, hence establishing its \emph{universality}. To obtain this result we extend the work of \cite{Meier_2012} by introducing an infinitary probability logic with truthful and non-epistemic conditioning events.

%% file: Text/Introduction/terminology_new.tex
\subsection{A Caveat on Terminology}
\label{subsec:terminology}

We want to emphasize one point about the terminology we use, namely our -- somewhat non-standard -- definition of universality. Starting from \cite{Mertens_Zamir_1985} and \cite{Heifetz_Samet_1998}, the notion of universal type structure has been associated with the idea that any other type structure can be uniquely embedded into the universal one.\footnote{There is an alternative notion of ``universality'' that can be found in the literature, for example in \cite{Brandenburger_Keisler_2006}, \cite{Friedenberg_2010}, and \cite{Fagin_et_al_1999}. According to this notion, a type structure is universal if there is an ordinal number $\alpha$ such that, for every ordinal $\beta > \alpha$, the $\beta$-belief order is the same as the $\alpha$-belief order, that is the $\alpha$-belief order determines all subsequent belief-orders (all the explicit constructions starting from a topological space satisfy this definition with $\alpha := \omega$, where $\omega$ is the ordinal counterpart of $\bN$). In \cite{Heifetz_Samet_1998b} this idea is used to prove that there is no universal (in the sense above) structure for knowledge spaces. See \cite{Aumann_1999a} for a  treatment of knowledge spaces.}

On the contrary, we consciously adopt an extension of the attempt of a standardization of the terminology on large type structures made in  \cite{Siniscalchi_2008}, where such taxonomy can be used for large \emph{preference} structures as well without modifications.\footnote{A preference structure is a structure that takes \emph{preferences} and not beliefs as primitive objects, building on the idea of \cite{Savage_1954} that beliefs can be derived from preferences. See \cite{Epstein_Wang_1996}, \cite{di_Tillio_2008} for explicit constructions of large preference structures with topological assumptions  and \cite{Ganguli_et_al_2016} for a topology-free construction.}  According to the small extension of the terminology of \cite{Siniscalchi_2008} we propose, a type structure is \emph{universal} if it is \emph{terminal}, \emph{belief-complete},\footnote{Belief-completeness has been explicitly introduced in the literature as \emph{completeness} by \cite{Brandenburger_2003}. \cite{Siniscalchi_2008} adopts that term (which should be the one used for preference structures), but we follow \cite{Meier_2012} in our terminology.} and \emph{non-redundant},\footnote{See \Sref{sec:non-redundancy} for the definition of non-redundancy, which is the notion that does not appear in the original taxonomy \cite{Siniscalchi_2008}. However, recent contributions have stressed the importance of this notion (e.g., \cite{Friedenberg_2010}).} where all these definitions are formally introduced in the course of the paper.

%% file: Text/Introduction/synopsis.tex
\subsection{Synopsis}

\Sref{sec:preliminaries} is devoted to introduce the mathematical concepts and notation used in the rest of the paper. In \Sref{sec:measure-theoretic_type_structure} we introduce type structures and the terminology we adopt. In \Sref{sec:terminality} we present a construction of the topology-free terminal type structure for conditional probability systems, which we show to be non-redundant in \Sref{sec:non-redundancy}. Finally, in \Sref{sec:universality} we construct a type structure, which we show to be the same as the one in \Sref{sec:terminality}, that we prove to be belief-complete, hence establishing its universality.

%% file: Text/preliminaries.tex
\section{Preliminaries}
\label{sec:preliminaries}

Given an arbitrary set $X$, we let $\wp ( X)$ denote the power set of $X$ and $\abs{X}$ its cardinality. Also, we let $\bQp := \bQ \cap [0 ,1]$. Recall that $\aleph_\gamma$ denotes an infinite cardinal number, where $\abs{ \bN} = \aleph_0$.\footnote{See \cite[Chapter 3]{Jech_2006} for an introduction to the topic.} We use the symbols ``$:=$'' in expressions of the form $X := \Set { \dots | \dots }$ and ``$\deff$'' in expressions of the form $X \deff Y$ with the meaning that the right-hand side \emph{defines} the left-hand side. Concerning logical symbols, we use $\neg$, $\wedge$, and $\vee$ to denote respectively ``not'', ``and'', and the inclusive reading of ``or''. We also employ the connective $\veebar$ to denote the \emph{exclusive} reading of the \emph{disjunction};\footnote{That is, ``$p \vee q$'' denotes the statement ``$p$ \emph{or} $q$ (or both)'' [inclusive disjunction] and ``$p \veebar q$'' denotes the statement: ``\emph{either} $p$ \emph{or} $q$ (but not both)'' [exclusive disjunction].} Finally, regarding the behavior of brackets, we adopt the usual conventions of eliminations according to decreasing priority with respect to $\neg$, $\bigwedge$, $\wedge$, $\bigvee$, $\vee$, $\veebar$, $\rightarrow$, $\leftrightarrow$. 

Let $(X , \Sigma_X)$ be a measurable space, that is, a set $X$ endowed with a $\sigma$-algebra $\Sigma_X$. If $X := \prod_{\lambda \in \Lambda} X_\lambda$ is an arbitrary product space, where every $(X_\lambda , \Sigma_\lambda)$ is a measurable space, then $X$ is endowed  with the  product $\sigma$-algebra induced by the $\sigma$-algebras of its component spaces, i.e., $\Sigma_{X}$ is the $\sigma$-algebra generated by sets of the form $\prod_{\lambda \in \Lambda} A_\lambda$, where $A_\lambda \in \Sigma_\lambda$ for every $\lambda \in \Lambda$ and $A_\lambda := X_\lambda$ except for finitely many $\lambda \in \Lambda$.

Let $\Delta (X)$ denote the set of all $\sigma$-additive probability measures over $X$ considered as a measurable space, endowed with the $\sigma$-algebra $\Sigma_{\Delta (X)}$ generated by all sets of the form
\begin{equation*}
\beta^{q} (A) = \Set { \mu \in \Delta (X) | \mu (A) \geq q },
\end{equation*}
where $A \in \Sigma_X$ and $q \in [0,1]$ or $q \in \bQp$.

Given $\Sigma_X$, fix a \emph{countable}\footnote{This requirement is not needed in \Sref{sec:terminality}, but we cannot dispense with it to establish the results in \Sref{sec:universality}.} subset $\mcal{B} \subseteq \Sigma_X \setminus \{ \varnothing \}$, and call the space $(X, \Sigma_X , \mcal{B})$ a \emph{conditional measurable space}.\footnote{To the best of our knowledge the term ``conditional measurable space'' has not been previously used in the literature. We adopt it because it seems a rather natural name for such an object.} 
The elements $B \in \mcal{B}$ can be considered as conditioning events. This gives rise to the following definition.

\begin{definition}[Conditional Probability System]
\label{def:CPS}
A \emph{conditional probability system} (henceforth, CPS) on a conditional measurable space $(X, \Sigma_X . \mcal{B})$ is a mapping
\begin{equation*}
\mu \Round { \cdot | \cdot } : \Sigma_X \times \mcal{B} \to [0,1]
\end{equation*}
that satisfies the following axioms:
\begin{enumerate}[label=A\arabic*., leftmargin=*, itemsep=0.5ex]
\item For all $B \in \mcal{B}$, $\mu (B | B) =1$.
\item For all $B \in \mcal{B}$, $\mu (\cdot | B)$ is a $\sigma$-additive probability measure on $(X, \Sigma_X)$.
\item For all $A \in \Sigma_X$, for all $B, C \in \mcal{B}$, if $A \subseteq B \subseteq C$, then $\mu (A | B) \ \mu(B|C) = \mu(A|C)$. 
\end{enumerate}
\end{definition}

\begin{notation}
For every $B \in \mcal{B}$, we let $\mu_B (\cdot) := \mu (\cdot | B)$.
\end{notation}

We constantly employ throughout the paper the functional-theoretical notation such that, given two arbitrary sets $X$ and $Y$, the set $Y^X$ denotes the family of all functions from $X$ to $Y$.

Let $[\Delta (X)]^{\mcal{B}}$ denote the set of all mappings from $\mcal{B}$ to $\Delta (X)$ and let $\Delta^\mcal{B} (X) \subseteq [\Delta (X)]^{\mcal{B}}$ denote the set of CPSs on $(X, \Sigma_X , \mcal{B})$, with typical elements $\mu = \big( \mu_B (\cdot) \big)_{B \in \mcal{B}} \in \Delta^\mcal{B} (X)$. That is, $\Delta^\mcal{B} (X)$ is the set of all mappings from $\mcal{B}$ to $\Delta (X)$ that satisfy conditions A1-A3 as in \Mref{def:CPS}. The set $\Delta^{\mcal{B}} (X)$ is endowed with the $\sigma$-algebra generated by sets of the form
\begin{equation*}
\beta^{p}_B (A) = \Set { \mu \in \Delta^{\mcal{B}} (X) | \mu_B (A) \geq p },
\end{equation*}
where $A \in \Sigma_X$, $B \in \mcal{B}$, and $p \in [0,1]$ or $p \in \bQp$.

\begin{notation}
We let $\Sigma_{\mcal{B} (X)}$ denote the $\sigma$-algebra on $\Delta^{\mcal{B}} (X)$ defined above.
\end{notation}

\begin{remark}
The space $(\Delta^{\mcal{B}} (X), \Sigma_{\mcal{B} (X)})$ is measurable.
\end{remark}

Let $(X, \Sigma_X)$ and $(Y, \Sigma_Y)$ be two measurable spaces and let $f \in Y^X$ be a  $(\Sigma_X, \Sigma_Y)$-measurable function.\footnote{Recall that a function $f \in Y^X$ is \emph{$(\Sigma_X, \Sigma_Y)$-measurable} if $f^{-1} (E) \in \Sigma_X$ for every $E \in \Sigma_Y$. In the following, when it is clear from the context, to lighten the text we omit the reference to the $\sigma$-algebras.} It is possible to define a sense in which these two measurable space are equal from a measurable perspective. This is captured by the following definition.

\begin{definition}[Measurable Isomorphism]
A \emph{measurable  isomorphism} between two measurable spaces $(X, \Sigma_X)$ and $(Y, \Sigma_Y)$ is a bijection $f \in Y^X$ such that both $f$ and $f^{-1}$ are measurable.
\end{definition}

If $(X, \Sigma_X)$ and $(Y, \Sigma_Y)$ are measurable spaces and $f \in Y^X$ is measurable, a $\sigma$-additive probability measure in $(\Delta (Y) , \Sigma_{\Delta (Y)})$ can induce a $\sigma$-additive probability measure in $(\Delta (X) , \Sigma_{\Delta (X)})$ via $f$.

\begin{definition}[Image Measure (Pushforward)]
\label{def:pushforward}
The \emph{image measure} (or the pushforward) of a $(\Sigma_X, \Sigma_Y)$-measurable function $f \in Y^X$ is  a $(\Sigma_{\Delta (X)}, \Sigma_{\Delta (Y)})$-measurable map $\widehat{f} : \Delta (X) \to \Delta (Y)$, such that 
\begin{equation*}
\widehat{f} (\mu) (E) := \mu (f^{-1} (E)),
\end{equation*}
for every $\mu \in \Delta (X)$ and for every $E \in \Sigma_Y$.
\end{definition}

In order to extend \Mref{def:pushforward} to the case of CPSs we introduce some additional notation. Let $(X, \Sigma_X, \mcal{B})$ be a conditional measurable space and define a product space $Z := X \times Y$, where $Y$ is an arbitrary measurable space endowed with a $\sigma$-algebra $\Sigma_Y$. Then, we define the family of conditioning events of $Z$ as
\begin{equation}
\label{eq:product_conditioning}
\mcal{B}_Z := \Set { C \subseteq Z | \exists B \in \mcal{B} : C := B \times Y }.
\end{equation}
By exploiting the structure of $\mcal{B}_Z$, we write $\Delta^{\mcal{B}} (Z)$ instead of $\Delta^{\mcal{B}_Z} (Z)$ and we extend \Mref{def:pushforward}.

\begin{definition}[Image Measure with Conditioning Events]
\label{def:pushforward_conditioning}
Let $(X, \Sigma_X, \mcal{B})$ be a conditional measurable space. Let  $Z := X \times Y$ and $Z' := X \times Y'$ be two product measurable spaces with $\sigma$-algebras $\Sigma_Z$ and $\Sigma_{Z'}$, and with $Y$ and $Y'$ arbitrary measurable spaces. Also, let $\mcal{B}$ the set of conditioning events of both $Z$ and $Z'$. Then, given a $(\Sigma_Z, \Sigma_{Z'})$-measurable function $f \in Z'^Z$,
the \emph{image measure with conditioning events}
$\widehat{f} :=  ( \widehat{f_B} )_{B \in \mcal{B}} : \Delta^{\mcal{B}} (Z) \to \Delta^{\mcal{B}} (Z')$ is defined by
\begin{equation*}
\widehat{f} (\mu_B) (E) := \mu_{B \times Y'} \Round { f^{-1} (E) }
\end{equation*}
for every CPS $\mu := (\mu_{B}) \in \Delta^{\mcal{B}} (Z)$ and for every $E \in \Sigma_{Z'}$.
\end{definition}

In the following, as it is customary, we let $I$ denote the set of players and $0$ stand for ``nature'', with $0 \notin I$. Then, we define $I_0 := I \cup \{ 0 \}$. We adopt the convention that we typically use $i$ for a representative element of $I_0$ and $j$ for a representative element of $I$. Also, given a family of sets $(X_i)_{i \in I_0}$, we let $X := \prod_{i \in I_0} X_i$ and $X_{-i} := \prod_{y \in I_0 \setminus \{i \}} X_y$ (the same convention applies to a family $(X_j)_{j \in I}$ modulo representative element).

\begin{definition}[Induced Function]
\label{def:induced}
Given a family of functions $(f_i)_{i \in I_0}$ of the form $f_i : X_i \to Y_i$, the \emph{induced function} $f : X \to Y$ is defined as
\begin{equation*} 
f \left( (x_i )_{i \in I_0} \right) := \left( f_i (x_i) \right)_{i \in I_0}.
\end{equation*}
\end{definition}

In the rest of the paper we will repeatedly use the following extension of \cite[Lemma 4.5]{Heifetz_Samet_1998} to address the case of conditional probabilities. As it is customary, given a set $X$ and an arbitrary family of subsets $\mcal{F} \subseteq 2^X$, we let $\sigma  (\mcal{F})$ denote the $\sigma$-algebra on $X$ generated by $\mcal{F}$.

\begin{lemma}
\label{lem:1}
Let $(X, \Sigma_X, \mcal{B})$ be a conditional measurable space. Let $\mcal{A}_X$ be an algebra such that $\Sigma_X := \sigma (\mcal{A}_X)$ and let $\mcal{A}_{\mcal{B} (X)}$ be the $\sigma$-algebra on $\Delta^{\mcal{B}} (X)$ generated by sets of the form 
\begin{equation*}
\Set { \beta^{p}_B (E) | E \in \mcal{A}_X , \ p \in [0,1], \ B \in \mcal{B} }. 
\end{equation*}
Then, $\mcal{A}_{\mcal{B} (X) } = \Sigma_{\mcal{B} (X)}$.  The same result holds if $p \in \bQp$.
\end{lemma}

Given an arbitrary product space $X := \prod_{\lambda \in \Lambda} X_\lambda$, we let $\proj_\lambda X$ denotes the projection on $X_\lambda$ of $X$, i.e., $\proj_\lambda : X \to X_\lambda$ is such that $\proj_\lambda (x) := x_\lambda$, where $x:= (x_\lambda)_{\lambda \in \lambda}$. As it is customary, given a measurable space $(X, \Sigma_X)$, for every $x \in X$ we let $\delta_x$ denote the Dirac measure on $(X, \Sigma_X)$: that is, for every $A \in \Sigma_X$, $\delta_{ x} (A)$ is the measure defined as $\delta_{x} (A) := 1$ if $x \in A $ and $\delta_{x} (A) := 0$ if $x \notin A$.\footnote{We indulge in the following abuse of notation, that is, for every $x \in X$, we write $\delta_x$ instead of $\delta_{\{ x \}}$.} Finally, given an arbitrary set $X$, we let $\Id_X$ denote the identity function on $X$, viz., the function $\Id_X : X \to X$ is such that $\Id_X (x) := x$.

%% file: Text/Type_Structures/intro.tex
\section{Type Structures}
\label{sec:measure-theoretic_type_structure}

For the rest of the paper we fix a measurable space $(S, \Sigma_S)$, where the set $S$ is the set of \emph{states of nature}, and a set of countable conditioning events $\mcal{B} \subseteq \Sigma_S \setminus \{ \varnothing \}$. Hence, we fix a conditional measurable space $( S, \Sigma_S, \mcal{B} )$. Observe that in referring to $S$ we use interchangeably the words ``parameter space'' and ``domain of uncertainty'', even if there is a conceptual difference between the two.\footnote{Indeed, ``parameter space'' is used in the incomplete information literature, while ``domain of uncertainty'' is more generic.}

In the following two sections we use two different typographical conventions to refer to the same objects. In \Sref{subsec:standard_type_structure} we use standard italicized \emph{serif} math symbols, while in \Sref{subsec:logical_type_structure} we use italicized \emph{sans-serif} math symbols.

%% file: Text/Type_Structures/standard.tex
\subsection{Standard Formulation}
\label{subsec:standard_type_structure}

First we provide a formal definition of what a type structure\footnote{Quite often authors use the word ``space'' to refer both to a tuple of objects such as $\mscr{T} := \la I_0 , ( S, \Sigma_S, \mcal{B} ), (T_i )_{i \in I_0 } , (m_j )_{j \in I} \ra$ \emph{and} the set of types $T_j$. We distinguish these objects by using  ``structure'' for the tuple and ``space'' for the set of types.} is on the conditional measurable space $( S, \Sigma_S, \mcal{B} )$.

\begin{definition}[Type Structure]
\label{def:type_structure}
A \emph{type structure} on a conditional measurable space $( S, \Sigma_S, \mcal{B} )$ is a tuple
\begin{equation*}\mscr{T} := \la I_0 , ( S, \Sigma_S, \mcal{B} ), (T_i )_{i \in I_0 } , (m_j )_{j \in I} \ra
\end{equation*}
of profiles of \emph{type spaces}, and \emph{belief functions} such that 
\begin{itemize}
\item[i)] $T_0 := S$ and $T_j$ are measurable spaces, called \emph{type spaces}, for every $j \in I$,
\item[ii)] $m_j := (m_{j, B})_{B \in \mcal{B}} : T_j \to \Delta^\mcal{B} (T)$ is a measurable function, called \emph{belief function}, for every $j \in I$,
\item[iii)] for every $j \in I$, for every $t_j \in T_j$, and for every $B \in \mcal{B}$, $\marg_{T_j} m_{j, B} (t_j) = \delta_{t_j}$.\footnote{This property, even if conceptually appealing, is not necessary for the construction (see for example \cite{Ganguli_et_al_2016}). \cite{Heifetz_Mongin_2001} and \cite{Meier_2012} distinguish type structures which possess this property from those which do not: in their terminology a type structure that satisifies this requirement is called an \emph{Harsanyi type space} (viz., structure).}
\end{itemize}
\end{definition}

An element $t \in T$ is called a \emph{state of the world}, with $T$ called the set of states of the world or \emph{state space}, while a $t_ i \in T_i$ is called an \emph{$i$-type}, for every $i \in I_0$. We have to distinguish the types that belong to nature, namely the elements of $T_0$, from the types of an arbitrary player $j \in I$, i.e., the elements of $T_j$. The $j$-types, for every $j \in I$, represent the epistemic states of player $j$.

\begin{remark}
Observe that point (ii) in \Mref{def:type_structure} is well-defined since $B \in \Sigma_S$, for every $B \in \mcal{B}$, and $\mcal{B}_T$ is equal to $\mcal{B}$ from \Mref{eq:product_conditioning}, since $T : = S \times \prod_{j \in I} T_j$.
\end{remark}


We modify for our context the adaptation of \cite{Battigalli_Siniscalchi_2002} of the $p$-belief operator of \cite{Monderer_Samet_1989} to address the presence of conditioning events. Hence, the event that individual $j \in I$ ascribes probability at least $p$ to an event $E \subseteq T$ given a conditioning event $B \in \mcal{B}$ is described by
\begin{equation}
\label{eq:p-belief}
\B^{p}_{j, B} (E) := \Set { t \in \proj^{-1}_j (t_j) |  m_{j, B} (t_j) (E) \geq p  }.
\end{equation}

Having formalized the notion of type structure, we provide a formal definition that captures when two type structures on a conditional measurable space $( S, \Sigma_S, \mcal{B} )$ can be regarded as being `equivalent'.

\begin{definition}[Type Morphism]
\label{def:type_morphism}
Take two type structures
\begin{equation*}
\mscr{T} := \la I_0 , (S, \Sigma_S, \mcal{B} ), (T_i )_{i \in I_0 } , (m_j )_{j \in I} \ra
\end{equation*}
and 
\begin{equation*}
\mscr{T}' := \la I_0 , (S, \Sigma_S, \mcal{B}) , (T'_i )_{i \in I_0 } , (m'_j )_{j \in I} \ra
\end{equation*} 
on the same conditional measurable space $(S, \Sigma_S, \mcal{B})$ and let $(f_i)_{i \in I_0}$ be an $I_0$-tuple of measurable functions $f_i : T_{i} \to T'_{i}$. The induced function $f :  T \to T'$ is called a \emph{type morphism} if
\begin{enumerate}
\item $f_0 := \Id_S$,
\item for every $j \in I$, $m'_j \circ f_j = \widehat{f} \circ m_j$, viz., the diagram
\begin{center}
\begin{tikzcd}
T_j \arrow[r, "f_j"] \arrow[d, "m_j"']
	& T'_j \arrow[d, "m'_j"] \\
\Delta^{\mcal{B}} (T) \arrow[r, "\widehat{f}"']
	& \Delta^{\mcal{B}} (T')
\end{tikzcd}
\end{center}
commutes.
\end{enumerate}
If $f$ is a measurable isomorphism, then the morphism is called a \emph{type isomorphism}.
\end{definition}

\begin{remark}
Observe that condition (2) is equivalent to saying that for every $j \in I$, $t_j \in T_j$, $B \in \mcal{B}$, and $E \subseteq T'$,
\begin{equation}
\label{eq:type_morphism}
m'_{j, B} ( f_{j} (t_j)) (E) = m_{j, B} ( t_j) ( f^{-1} (E) ).
\end{equation}
Moreover, $f$ preserves belief operators, i.e.,
$$ \B^{p}_{j, B} ( f^{-1} (E) ) = f^{-1} ( \B^{p}_{j, B} (E) ),$$
for every $0 \leq p \leq 1$, $j \in I$, $E \subseteq T'$, and $B \in \mcal{B}$.
\end{remark}

%% file: Text/Type_Structures/logic.tex
\subsection{Logical Reformulation}
\label{subsec:logical_type_structure}

We can provide a more refined description of what $S$ actually is. Let $\mathsf{X}$ be a set of primitive propositions with typical element $\mathsf{p}$. For every $\mathsf{p} \in \mathsf{X}$, we add the negation of $\mathsf{p}$, that is, $\neg \mathsf{p}$, thus obtaining $\otbar{\mathsf{X}}$, with typical element $\vf$. 
We now define an \emph{exogenously} imposed family of sets of \emph{conditioning propositions} $\Bp$ as follows:
\begin{equation*}
\Bp := \Set { \F \in \wp ( \otbar{\mathsf{X}}) \setminus \{\varnothing\}  | %
\forall \vf \in \otbar{\mathsf{X}} \ ( \vf \in \F \Rightarrow \neg \vf \notin \F ) } 
\end{equation*}
Again, we denote a typical element of a set of conditioning propositions $\F$ with $\vf$. 
Observe that this definition ensures that every $\F \in \Bp$ is not empty and that  $\Bp$ is comprised of set of propositions which are \emph{consistent}, that is, which do not contain both $\mathsf{p}$ and $\neg \mathsf{p}$, for every $\mathsf{p} \in \otbar{\mathsf{X}}$. 

From the tuple $(\Xp , \Bp)$ we can always retrieve a conditional measurable space $(S , \Sigma_S , \mcal{B})$ as follows.\footnote{Observe that, even if $\Bp$ is defined in terms of $\otbar{\mathsf{X}}$, we still consider $\Xp$ as our primitive object.} The set of states of nature $S$ can be defined as 
\begin{equation*}
S := \Set { s | \forall \vf \in \otbar{\mathsf{X}} \ ( \vf \in s \veebar \neg \vf \in s  }.\footnote{As pointed out in \cite{Meier_2012}, $S$ could be simply defined as $S := \wp (X)$. We are taking this somewhat longer route of enlarging $\Xp$ to $\otbar{\mathsf{X}}$, because we have to define an additional obejct, namely $\Bp$.}
\end{equation*}
Hence, a state of nature $s \in S$ is list of primitive propositions that is complete and consistent: it is \emph{complete} since for every $\mathsf{p} \in \Xp$ there is an occurence of either $\mathsf{p}$ or  $\neg \mathsf{p}$, and it is \emph{consistent} since the previous ``or'' has to be read in its exclusive meaning. We endow $S$ it with the $\sigma$-algebra $\Sigma_S$ defined as
\begin{equation*}
\Sigma_S := \sigma \big( \Set { s \in S | \forall \vf \in \otbar{\mathsf{X}} \ %
( \vf \in s \veebar \neg \vf \in s  } \big) ,
\end{equation*} 
where $\vf$ denotes either a primitive proposition $\mathsf{p}$ or its negation $\neg \mathsf{p}$.
Finally, $\mcal{B}$ is defined as
\begin{equation*}
\mcal{B} := \Set { B \in \Sigma_S \setminus \{ \varnothing \} | %
\exists \F \in \Bp : 
\begin{array}{l}
\ i) \ \ \forall \vf  \in \Phi \ \forall s \in B \ (\vf  \in s ), \\%
\ ii) \  \forall \mathsf{p} \in \Xp \ ( \mathsf{p} \notin \F, \neg \mathsf{p} \notin \F \Longrightarrow \exists s, s' \in B : \mathsf{p} \in s,  \neg \mathsf{p} \in s' )
\end{array}
}.
\end{equation*}
Hence,  every state $s \in B$, for every $B \in \mcal{B}$, is again a complete and consistent list of propositions from $\otbar{\mathsf{X}}$. 
Before introducing the first definition of this section, we introduce a new bit of notation. We let $\top$ denote a tautology and we adopt the convention that $\mathsf{X}_\top := \mathsf{X} \cup \{ \top \}$. We are now in position to define type structures for this setting, which provides us a more fine grained perception of the objects under scrutiny.

\begin{definition}[Type Structure with Valuation Function]
\label{def:propositional_type_structure}
Fix  a tuple $(\Xp , \Bp)$, which induces a conditional measurable space $(S, \Sigma_S , \mcal{B})$. A \emph{type structure} on $(\Xp , \Bp)$ is a tuple
\begin{equation*}
\mbf{T} := \la I_0 , ( \Xp_\top, \Bp ), (S, \Sigma_S , \mcal{B}) , (\T_i )_{i \in I_0 } , (\mathsf{m}_j )_{j \in I} , \val \ra
\end{equation*}
of profiles of \emph{type spaces}, and \emph{belief functions} such that 
\begin{itemize}
\item[i)] $\T_{0} :=  S$ and $\T_j $ are measurable spaces, called \emph{type spaces}, where $\T_j$ is defined for every $j \in I$, and with $\T := \prod_{i \in I_0} \T_i$;
\item[ii)] $\m_j := (\m_{j, B})_{B \in \mcal{B}} : \T_j \to \Delta^\mcal{B} (\T)$ is a measurable function, called \emph{belief function}, for every $j \in I$;
\item[iii)] for every $j \in I$, for every $\ttt_j \in \T_j$, and for every $B \in \mcal{B}$, $\marg_{\T_j} \m_{j, B} (\ttt_j) = \delta_{\ttt_j}$;
\item[iv)] $\val : S \times \Xp_\top \to \Set { 0,1}$ is a measurable function, called \emph{valuation function} such that 
\begin{itemize}[leftmargin=*]
\item for every $\mathsf{p} \in \Xp_\top$
\begin{equation*}
\val \Round { \cdot , \mathsf{p} }:=
\begin{cases}
1, & \text{ if } \mathsf{p} \in s,\\
0, & \text{ if } \mathsf{p} \notin s,
\end{cases}
\end{equation*}
\item $\val ( s, \top ) =1$ for every $s \in S$.
\end{itemize}
\end{itemize}
\end{definition}


Much in the same spirit of the previous section, we want to be able to say if two type structures on the same domain of uncertainty are the same in this framework as well. The following definition captures this.

\begin{definition}[Type Morphism with Valuation Function]
\label{def:type_morphism_valuation}
Take two type structures
\begin{equation*}
\mbf{T} := \la I_0 , ( \Xp_\top , \Bp ), (S, \Sigma_S , \mcal{B}) , (\T_i )_{i \in I_0 } , (\m_j )_{j \in I} , \val \ra
\end{equation*}
and 
\begin{equation*}
\mbf{T}' := \la I_0 , ( \Xp_\top , \Bp ), (S, \Sigma_S , \mcal{B}) , (\T'_i )_{i \in I_0 } , (\m'_j )_{j \in I} , \val \ra
\end{equation*} 
on the same conditional measurable space $(S, \Sigma_S, \mcal{B})$ and let $(\tens{f}_i)_{i \in I_0}$ be an $I_0$-tuple of measurable functions $\tens{f}_i : \T_{i} \to \T'_{i}$. The induced function $\tens{f} :  \T \to \T'$ is called a \emph{type morphism} if
\begin{enumerate}
\item $\tens{f}_0 := \Id_S$,
\item for every $j \in I$, $\m'_j \circ \tens{f}_j = \widehat{\tens{f}} \circ \m_j$, viz., the diagram
\begin{center}
\begin{tikzcd}
\T_j \arrow[r, "\tens{f}_j"] \arrow[d, "\m_j"']
	& \T'_j \arrow[d, "\m'_j"] \\
\Delta^{\mcal{B}} (T) \arrow[r, "\widehat{\tens{f}}"']
	& \Delta^{\mcal{B}} (\T')
\end{tikzcd}
\end{center}
commutes,
\item for every $s \in S$ and for every $\mathsf{p} \in \Xp_\top$
\begin{equation*}
\val ( s , \mathsf{p} ) = \val ( \tens{f}_0 ( s ) , \mathsf{p} ).
\end{equation*}
\end{enumerate}
If $\tens{f}$ is a measurable isomorphism, then the morphism is called a \emph{type isomorphism}.
\end{definition}

%% file: Text/Type_Structures/large.tex
\subsection{Large Type Structures}
\label{subsec:large_type_structures}

In this section we collect the definitions of large type structures that we employ in this paper. Observe that, as pointed out in \Sref{subsec:terminology}, they are all standard with one exception, namely the definition of universality. First we introduce a notion that is going to be crucial in the remainder of the paper.

\begin{definition}[Class of Type Structures]
\label{def:type_class}
We let $\mfrak{T}$ denote the class of all type structures $\mscr{T}$ on $(S, \Sigma_S , \mcal{B})$ with set of players $I$. 
\end{definition}

In the following definition, by employing the notation introduced in \Sref{subsec:standard_type_structure}, we do not distinguish between type structures with and without valuation functions.

\begin{definition}[Belief-Complete Type Structure]
\label{def:belief-complete_type_structure}
A type structure $\splitatcommas{\otbar{\mscr{T}} := \la I_0 , ( S, \Sigma_S, \mcal{B}) , (\otbar{T}_i )_{i \in I_0 } , (\otbar{m}_j )_{j \in I} \ra}$  in $\mfrak{T}$ is \emph{belief-complete} if, for every $j \in I$, the function $\otbar{m}_j$ is surjective.
\end{definition}

\begin{definition}[Terminal Type Structure]
\label{def:universal_type_structure}
A type structure $\otbar{\mscr{T}} := \la I_0 , ( S, \Sigma_S, \mcal{B}) , (\otbar{T}_i )_{i \in I_0 } , (\otbar{m}_j )_{j \in I} \ra$ in $\mfrak{T}$ is \emph{terminal} if for every other type structure $\mscr{T}$ in $\mfrak{T}$ there is a unique type morphism from $\mscr{T}$ to $\otbar{\mscr{T}}$.\footnote{Observe that this definition can be translated in category-theoretical terms by saying that $\otbar{\mscr{T}}$ is terminal in the category of type structures on $(S, \Sigma_S , \mcal{B})$. Not surprisingly, this is the origin behind the usage of this attribute for such large type structures, which was advocated by \cite{Armbruster_Boge_1979} and \cite{Boge_Eisele_1979}. Other papers that explicitly refer to a categorical reformulation of the problem at hand are \cite{Vassilakis_1991}, \cite{Vassilakis_1992}, and \cite{Pinter_2010}. Alternatively, $\otbar{\mscr{T}}$ can be seen as a terminal coalgebra (see \cite{Rutten_2000} for an introduction to coalgebras). Building on previous work by \cite{Viglizzo_2005} and \cite{Moss_Viglizzo_2006}, this is the path chosen by \cite{Heinsalu_2013} to construct the topology-free terminal type structure with unawareness. An alternative definition of terminality for topological settings can be found in \cite{Friedenberg_2010}, where a terminal type structure $\otbar{\mscr{T}}$ is defined as a type structure such that, for every player $j \in I$, for every type $t_j \in T_j$ in an arbitrary type structure $\mscr{T}$, there is a type $\otbar{t}_j \in \otbar{T}_j$ in $\otbar{T}$ such that $t_j$ and $\otbar{t}_j$ induce the same coherent hierarchy of beliefs.}
\end{definition}

Since \emph{non-redundancy} needs a specific apparatus that we develop in \Sref{sec:terminality}, we postpone the formal definition of this notion until \Sref{sec:non-redundancy}. As mentioned in \Sref{subsec:terminology}, the definition of \emph{universality} we use  that comes next is not standard, but it is in line with the terminology introduced in \cite{Siniscalchi_2008}.

\begin{definition}[Universal Type Structure]
\label{def:uni_type_structure}
A type structure $\otbar{\mscr{T}} := \la I_0 , ( S, \Sigma_S, \mcal{B}) , (\otbar{T}_i )_{i \in I_0 } , (\otbar{m}_j )_{j \in I} \ra$ is \emph{universal} if it is belief-complete, terminal, and non-redundant (as in \Mref{def:non-redundancy}).
\end{definition}

The goal of this paper is to show that, for every conditional measurable space $(S , \Sigma_S , \mcal{B})$, there is a type structure that is universal as in \Mref{def:uni_type_structure}.

%% file: Text/Terminality/terminality.tex
\section{Terminality}
\label{sec:terminality}

In this section we provide a proof of the following theorem. 

\begin{theorem}
\label{th:main_theorem}
For every conditional measurable space $( S, \Sigma_S, \mcal{B} )$ there exists a terminal type structure $\splitatcommas{\mscr{T}^* := \la I_0 , ( S, \Sigma_S, \mcal{B}) , (T^{*}_i )_{i \in I_0 } , (m^{*}_j )_{j \in I} \ra}$ on $( S, \Sigma_S, \mcal{B} )$ that is unique up to measurable isomorphism.
\end{theorem}

\begin{remark}
\label{rem:uniqueness_terminality}
\cite[Proposition 3.5]{Heifetz_Samet_1998} prove that there is at most one terminal type structure on a measurable space $(S, \Sigma_S)$ up to measurable isomorphism. Their proof applies to our case of a conditional measurable space $( S, \Sigma_S, \mcal{B} )$ without modifications.
\end{remark}

The next two sections, namely \Sref{subsec:theoretical_framework_hierarchies} and \Sref{subsec:terminal_type_infinite}, are devoted to prove this theorem via what are called infinite hierarchies of beliefs.

%% file: Text/Terminality/infinite_hierarchies.tex
\subsection{Infinite Hierarchies of Beliefs -- Theoretical Framework}
\label{subsec:theoretical_framework_hierarchies}

We let $\mscr{T} := \la I_0 , ( S, \Sigma_S, \mcal{B} ), (T_i )_{i \in I_0 } , (m_j )_{i \in I} \ra$ be an arbitrary type structure on the conditional measurable space $(S, \Sigma_S, \mcal{B} )$, where $S$ denotes the set of states of nature.

The idea behind this construction is to build infinite hierarchies of beliefs, i.e., beliefs of increasing order, and, for every belief order, a corresponding family of conditioning events.

\begin{definition}[Infinite Hierarchies of Beliefs]
For every $k \geq 0$, let $H^{k}_0 := S$. For every $j \in I$, $H^{0}_j$ is a singleton. Then, proceed with the following inductive construction:
\begin{center}
\begin{tabular}{cc}
$ H^{0}_{j} := \Set { \theta_j }$, 		&%
$ \mcal{B}^{0} :=  \mcal{B} $, \\
$\vdots$			& 	$\vdots$ \\
$ H^{k+1}_{j} = H^{k}_{j} \times \Delta^{\mcal{B}^{k}} (H^{k} ) $, &%
$\mcal{B}^{k+1} = \Set { C \subseteq H^{k+1}_j | \exists B \in \mcal{B}^{k} : C = B \times \Delta^{\mcal{B}^{k}} (H^{k}) } $,   \\
$\vdots$		&	 	$\vdots$ \\
\end{tabular}
\end{center}
The space of $j$-hierachies for player $j$ is 
\begin{equation}
\label{eq:j-hierarchies}
H^{\mcal{B}}_{j} :=  H^{0}_{j} \times  \prod_{l = 0}^{\infty} \Delta^\mcal{B} (H^l) ,
\end{equation}
where $H^k := \prod_{i \in I_0 } H^{k}_i$ for every $k \geq 0$. By having for nature $H_0 := S$, let $H^{\mcal{B}} := H_0 \times \prod_{j \in I } H^{\mcal{B}}_j $. The space $H^{\mcal{B}}$ is called the \emph{hierarchies space}.
\end{definition}

\begin{notation}
Observe that in the previous definition and in what follows the symbol $\mcal{B}$ in conjunction with the hierarchies space, e.g., the expression ``$H^{\mcal{B}}_i$'' in \Mref{eq:j-hierarchies}, acts just as a reminder that we are dealing with conditioning events and should not be read in the functional analytic way.
\end{notation}

Thus, for every $k \geq 1$, $\mcal{B}^{k+1} = \mcal{C} (\mcal{B}^k) := \Set { C \subseteq H^{k+1}_j | \exists B \in \mcal{B}^{k} : C = B \times \Delta^{\mcal{B}^{k}} (H^{k}_j) }$ represents an original conditioning event $B \in \mcal{B}^0$ to the $k$-order belief (indeed, recall that $\mcal{B}^0 := \mcal{B}$). For this reason, in \Mref{eq:j-hierarchies} we write $\mcal{B}$, without mentioning the hierarchy of conditioning events: that is, for every $k$ we have $\Delta^{\mcal{B}^k} (H^{k}_j) = \Delta^{\mcal{B}} (H^{k}_j)$.

We define $\pi^{k}_i : H^{\mcal{B}}_i \to H^{k}_i$ as a projection function on hierarchies, with the induced map defined as $\pi^k  : H^{\mcal{B}} \to H^k$. This is a crucial tool in the following definition, which is needed to unpack information, for every conditioning event $B \in \mcal{B}$, from the spaces $T_i$, for every $i \in I_0$.

\begin{definition}[$i$-Description Map]
\label{def:i-description_map}
The function $h_i := (h_{i, B})_{i \in I_0, B \in \mcal{B}} : T_i \to H^{\mcal{B}}_i$ is called the \emph{$i$-description map}, for every $i \in I_0$, and it is inductively defined as follows:
\begin{itemize}[leftmargin=*]
\item for $i=0$, and for every $k\geq 0$, $h^{k}_0 := \Id_S$;
\item for $j \in I$, $h^{0}_{j, B}$ is uniquely defined since $H^{0}_j$ is a singleton, while $h^{k+1}_{j, B}$, for every $B \in \mcal{B}$, is defined as 
\begin{equation*}
h^{k+1}_{j, B} (t_j) \big(  h^{k}_{j, B} (t_j), m_{j, B} (t_j) \circ ( h^{k}_{B})^{-1} \big) := %
\left(  h^{0}_{j, B} (t_j) , m_{j, B} (t_j) \circ ( h^{0}_{B} )^{-1} , \dots ,  m_{j, B} (t_j) \circ ( h^{k}_{B} )^{-1} \right),
\end{equation*}
where $h^{k}_B := (h^{k}_{i, B})_{i \in I_0}$.
\end{itemize}
Then, for every $i \in I_0$, the \emph{$i$-description of $t_i$ at $B \in \mcal{B}$} is the element $h_{i, B} (t_i)$ defined as the unique function from $T_i$ to $H_i$ such that $h^{k}_{i, B} = \pi^{k}_{i} \circ h_{i, B}$, for every $k \geq 0$, i.e., 
\begin{equation*}
h_{i, B} (t_i) := %
\left(  h^{0}_{i, B} (t_i) , m_{i, B} (t_i) \circ ( h^{0}_{B} )^{-1} , \dots ,  m_{i, B} (t_i) \circ ( h^{k}_{B} )^{-1}, \dots \right) . 
\end{equation*}
Finally, let $h_0 := \Id_S$.
\end{definition}

\begin{definition}[Description Map]
The \emph{description map} is the unique induced function
\begin{equation*}
h := (h_{i, B} )_{ i \in I_0, B \in \mcal{B}} : T \to H^{\mcal{B}}.
\end{equation*}
The element $h (t)$ is called the \emph{description} of $t$.
\end{definition}

We want to be sure that hierarchies are well-behaved under type morphisms,something which we establish next.

\begin{proposition}
\label{prop:morphisms_descriptions_hierarchies}
Type morphisms preserve descriptions and $i$-descriptions.
\end{proposition}

\subsection{The Terminal Type Structure via Infinite Hierarchies of Beliefs}
\label{subsec:terminal_type_infinite}

The construction of the terminal type structure $\mscr{T}^*$ comes in various steps. Ffirst of all we want to define $\mscr{T}^*$ and then we want to show that it is a type structure as in \Mref{def:type_structure}. To do so, we want to construct a profile of type spaces $(T^{*}_i)_{i \in I_0}$ that are measurable and a profile of belief functions $(m^{*}_j)_{j \in I}$ that are measurable.

We start from the type spaces. In doing so, we do not actually construct the types. Rather we explicitly assume the existence of type structures on the conditional measurable space $(S, \Sigma_S, \mcal{B})$ and we construct the terminal type spaces as objects that contain all the possible types living in all the possible type structures on $(S, \Sigma_S, \mcal{B})$.

\begin{definition}[The Terminal Type Spaces for Infinite Hierarchies]
Let $( S, \Sigma_S, \mcal{B})$ be a conditional measurable space and for every $i \in I_0$ define type spaces $T^{*}_i$ on $( S, \Sigma_S, \mcal{B} )$ as follows:
\begin{itemize}[leftmargin=*]
\item $T^{*}_0 := S$,
\item for every $j \in I$
\begin{equation*}
T^{*}_j := \Set { t^{*}_j \in H_j | \exists t_j \in T_{j} : t^{*}_j = (h_{j, B} (t_j))_{B \in \mcal{B}} },
\end{equation*}
where $T_j$ is a type space from a type structure $\mscr{T}$ in the class of type structures $\mfrak{T}$ on $( S, \Sigma_S, \mcal{B} )$.
\end{itemize}
Finally, endow $T^{*}_i$, with the $\sigma$-algebra inherited from $H_i$.
\end{definition}

\begin{definition}
\label{def:belief_function}
For every $j \in I$, let $m^{*}_j := (m^{*}_{j, B})_{j \in I, B \in \mcal{B}} : T^{*}_j \to \Delta^\mcal{B} ( T^{*})$ be a function defined by
\begin{equation}
\label{eq:belief_function}
m^{*}_{j, B} ( t^{*}_{j} ) = m_{j, B} (t_j)  \circ h^{-1}_B ,
\end{equation} 
for every $t_j \in T_j$ and for every $B \in \mcal{B}$.
\end{definition}

\begin{remark}
Observe that $m^{*}_j$ is a $\sigma$-additive probability measure.
\end{remark}

Observe now that, in proving the following theorem, not only we have to prove that the function defined as in \Mref{def:belief_function} is measurable for every $j \in I$, but we also have to prove that it is actually a belief function as in \Mref{def:type_structure}. 

\begin{theorem}
\label{th:type_structure_hierarchies}
The tuple $\mscr{T}^* := \la I_0 , ( S, \Sigma_S, \mcal{B}) , (T^{*}_i )_{i \in I_0 } , (m^{*}_j )_{j \in I} \ra$ is a type structure on $( S, \Sigma_S, \mcal{B})$.
\end{theorem}

The next theorem establishes the result that the type structure $\mscr{T}^*$ defined above is the terminal type structure, hence proving  \Mref{th:main_theorem}.

\begin{theorem}[Terminality of $\mscr{T}^*$ via hierarchies]
\label{th:universal_hierarchies}
The type structure $$\mscr{T}^* := \la I_0 ,( S, \Sigma_S, \mcal{B}), (T^{*}_i )_{i \in I_0 } , (m^{*}_j )_{j \in I} \ra$$ is the terminal type structure on $(S, \Sigma_S, \mcal{B})$.
\end{theorem}

%% file: Text/non-redundancy.tex
\section{Non-redundancy}
\label{sec:non-redundancy}

In this section we adapt the definition of non-redundancy in \cite{Liu_2009} to the construction performed in \Sref{sec:terminality}.\footnote{See \cite[Definition 6]{Liu_2009}, \cite[Definition 2.5]{Mertens_Zamir_1985}, and \cite{Friedenberg_Meier_2011}.} We let $\sigma (h_i)$ denote the smallest $\sigma$-algebra over $T_i$ for which the $i$-description map $h_i$ defined in \Mref{def:i-description_map} is measurable, for every $i \in I_0$.

\begin{definition}[Non-redundancy]
\label{def:non-redundancy}
A type structure $\mscr{T} := \la I_0 , (S, \Sigma_S, \mcal{B} ), (T_i )_{i \in I_0 } , (m_j )_{j \in I} \ra$ in $\mfrak{T}$ \emph{non-redundant} if, for every $j \in I$, $\sigma (h_j)$ is a $\sigma$-algebra that separates points in $T_j$.\footnote{A $\sigma$-algebra $\Sigma_X$ on $X$ separates point if for every $x, x' \in X$ there exists a $E \in \Sigma_X$ such that $x \in E$ and $x' \notin E$.}
\end{definition}

Non-redundancy coincides with the idea that the hierarchy maps are injective. \cite[Proposition 2]{Liu_2009} characterizes non-redundancy and states that the topology-free terminal type structure of \cite{Heifetz_Samet_1998} is non-redundant. Here, by adapting it to the presence of conditioning events, we state the proposition dividing it in two parts.

\begin{proposition}[Characterization of non-redundancy {\cite[Proposition 2]{Liu_2009}}]
\hspace{1cm}
\begin{enumerate}[leftmargin=*]
\item A type structure $\mscr{T} := \la I_0 , (S, \Sigma_S, \mcal{B} ), (T_i )_{i \in I_0 } , (m_j )_{j \in I} \ra$ is non-redundant if and only if the hierarchy map $h_j : T_j \to H^{\mcal{B}}_j$ is injective.
\item A non-redundant type structure separates point as in \Mref{def:non-redundancy}.
\end{enumerate}
\end{proposition}

The second part of the proposition, that corresponds to point (3) of Proposition 2 in \cite{Liu_2009}, establishes the non-redundancy of our type structure $\mscr{T}^* := \la I_0 , ( S, \Sigma_S, \mcal{B}) , (T^{*}_i )_{i \in I_0 } , (m^{*}_j )_{j \in I} \ra$ constructed in \Sref{subsec:terminal_type_infinite}.

\begin{theorem}[Non-redundancy of $\mscr{T}^*$ {\cite[Proposition 2]{Liu_2009}}]
\label{th:non-redundancy}
The terminal type structure $\splitatcommas{\mscr{T}^* := \la I_0 , ( S, \Sigma_S, \mcal{B}) , (T^{*}_i )_{i \in I_0 } , (m^{*}_j )_{j \in I} \ra}$ is non-redundant.
\end{theorem}

%% file: Text/Belief-Completeness/intro.tex
\section{Belief-Completeness}
\label{sec:universality}

In this section we prove that the type structure $\mscr{T}^* := \la I_0 , ( S, \Sigma_S, \mcal{B}) , (T^{*}_i )_{i \in I_0 } , (m^{*}_j )_{j \in I} \ra$ previously constructed  in \Sref{subsec:terminal_type_infinite}, which we have shown to be terminal and non-redundant, is also belief-complete as in \Mref{def:belief-complete_type_structure}.

%% file: Text/Belief-Completeness/infinitary_logic/intro.tex
\subsection{Infinitary Logic for Type Structures}
\label{subsec:infinitary_logic}

In this section we introduce our \emph{language}, which is  an opportune modification of the infinitary probabilistic logic for type structures introduced in \cite{Meier_2012} to deal with updating beliefs.

%% file: Text/Belief-Completeness/infinitary_logic/syntax.tex
\subsubsection{Syntax}
\label{subsubsec:syntax}

Recall from \Sref{subsec:logical_type_structure} that a tuple $(\Xp , \Bp)$ always induces a conditional measurable space $(S , \Sigma_S , \mcal{B})$. However, again from \Sref{subsec:logical_type_structure}, the original domain of uncertainty is $(\Xp_\top , \Bp)$, that is, the set of primitive propositions $\Xp$ augmented with the symbol $\top$, which denotes the constant truth, and the family of set of conditioning propositions $\Bp$ induced from $\Xp$ via $\otbar{\mathsf{X}}$.

\begin{definition}[Finitary Formulae]
\label{def:finitary_formulae}
The set $\Fin$ of \emph{finitary formulae} is the least set such that:
\begin{itemize}[leftmargin=*]
\item $\mathsf{p} \in \Fin$, for every $\mathsf{p} \in \Xp_\top$;
\item if $\vf \in \Fin$, then $\neg \vf \in \Fin$;
\item if $\vf, \psi \in \Fin$, then $\vf \wedge \psi \in \Fin$;
\item if $\vf \in \Fin$, then $\p^{\alpha}_{j , \F} (\vf) \in \Fin$, for every $j \in I$, $\F \in \Bp$, and $\alpha \in \bQp$.
\end{itemize}
\end{definition}

\begin{remark}
The operator $\p^{\alpha}_{j , \F} (\vf)$  captures the statement ``given information $\F$ individual $j$ assigns probability \emph{at least} $\alpha$ to $\vf$''
\end{remark}

\begin{definition}[Cardinality of $\Fin$]
The cardinality of $\Fin$ is defined as $\abs{\Fin} : = \max \Set { \abs{I}, \abs{\mathsf{X}}, \abs{\aleph_0} }$ with $\abs{\Fin} = \aleph_\gamma$ for a cardinal number $\gamma$.
\end{definition}

From \Mref{def:finitary_formulae} we construct the infinitary language that we adopt in this paper.

\pagebreak[3]

\begin{definition}[Formulae]
\label{def:formulae}
The set $\mcal{L}$ of \emph{formulae} is the least set such that:
\begin{itemize}[leftmargin=*]
\item $\varphi \in \mcal{L}$, for every $\varphi \in \Fin$;
\item if $\varphi \in \mcal{L}$, then $\neg \varphi \in \mcal{L}$;
\item if $\Gamma \subseteq \mcal{L}$ such that $\abs{\Gamma} \leq 2^{\aleph_\gamma}$, then $\big( \bigwedge_{\varphi \in \Gamma} \varphi \big) \in \mcal{L}$.
\end{itemize}
\end{definition}

Observe that the set of formulae $\mcal{L}$ can be seen as comprised of two different parts: one part deals with the statements concerning nature, while the other with all those statements that pertain to an individual $j$, for every $j \in I$. the following two definitions capture this intuition.

\begin{definition}[$0$-Formulae]
The set $\mcal{L}_0$ of \emph{formulae} is the set of (infinitary) propositional formulae in $\mcal{L}$, where the infinitary part comes from the last condition in \Mref{def:formulae}.
\end{definition}

\begin{definition}[$j$-Formulae]
For every $j \in I$ the set $\mcal{L}_j$ of $j$-\emph{formulae} is the least set of formulae such that:
\begin{itemize}[leftmargin=*]
\item if $\varphi \in \Fin$, then $\mbf{p}^{\alpha}_{j , \F} (\varphi) \in \mcal{L}_j$, for every $\F \in \Bp$ and $\alpha \in \bQp$;
\item if $\varphi \in \mcal{L}_j$, then $\neg \varphi \in \mcal{L}_j$;
\item if $\Phi \subseteq \mcal{L}_j$ such that $\abs{\Phi} \leq 2^{\aleph_\gamma}$, then $\big( \bigwedge_{\varphi \in \Phi} \varphi \big) \in \mcal{L}_j$.
\end{itemize}
\end{definition}

\begin{notation}
We let $\otbar{\mcal{L}}_{i} := \Fin \cap \mcal{L}_i$, for every $i \in I_0$.
\end{notation}

%% file: Text/Belief-Completeness/infinitary_logic/semantics.tex
\subsubsection{Semantics}
\label{subsubsec:semantics}

Our starting point for the semantics of our infinitary language is \Mref{def:propositional_type_structure}, i.e., the definition of type structure with valuation function.

\begin{notation}
Fix a type structure $\mbf{T} := \la I_0 ,  ( \Xp_\top, \Bp ), ( S, \Sigma_S, \mcal{B} ), (\T_i )_{i \in I_0 } , (\m_j )_{j \in I}, \val \ra$. Then, for every $B \in \mcal{B}$, we let 
\begin{equation*}
[ \m_{j, B} (\ttt_j) ] := \Set { (\proj^{-1} (\ttt'_j) ) \in \T | \m_{j , B} (\ttt_j) = \m_{j, B} (\ttt'_j) } 
\end{equation*}
for every $j \in I$.  
\end{notation}

\begin{remark}[Product Type Structure]
Since all the definitions we have provided in \Sref{sec:measure-theoretic_type_structure} of a type structure, irrespective of the presence or not of the valuation function, correspond to a \emph{product} type structure, it is understood that in the following $\ttt : = (\ttt_i )_{i \in I_0} = (s , (\ttt_j)_{j \in I} )$. 
\end{remark}

\begin{notation}
For every $\vf \in \Fin$, we let $\evaluation[\T ]{\varphi} := \Set { \ttt \in \T | (\mbf{T} , \ttt ) \models \varphi }$.
\end{notation}

\begin{definition}[Model for Type Structure]
Fix a type structure $\mbf{T}$ on $(\Xp_\top , \Bp)$. Then:
\begin{itemize}[leftmargin=*]
\item $(\mbf{T} , \ttt ) \models \top$ always; 
\item for every $\mathsf{p} \in \Xp_\top$, 
\begin{equation*}
(\mbf{T} , \ttt ) \models \mathsf{p} \deff \val (s , \mathsf{p}) = 1;
\end{equation*} 
\item for every $\vf, \psi \in \mcal{L}$, 
\begin{equation*}
(\mbf{T} , \ttt ) \models \varphi \wedge \psi \deff (\mbf{T} , \ttt ) \models \vf , (\mbf{T} , \ttt ) \models \psi ;
\end{equation*} 
\item for every $\vf \in \mcal{L}$, 
\begin{equation*}
(\mbf{T} , \ttt) \models \neg \vf \deff (\mbf{T} , \ttt) \not\models \vf ;
\end{equation*} 
\item for every $j \in I$, for every $\alpha \in \bQp$, for every $\vf \in \Fin$ such that $\evaluation[\T]{\vf} \in \Sigma_S$, and for every $\Phi \in \Bp$ with corresponding $B \in \mcal{B}$ 
\begin{equation*}
(\mbf{T} , \ttt ) \models \p^{\alpha}_{j, \Phi} ( \varphi) \deff \m_{j, B} ( \ttt_j ) \Big( \evaluation[\T]{\vf} \Big) \geq \alpha ,
\end{equation*} 
where $\ttt_j = \proj_j \ttt$.
\end{itemize}
\end{definition}

\begin{definition}[Valid Formula]
A formula $\vf \in \mcal{L}$ is \emph{valid} in the class of type structures $\mfrak{T}$ over $(\Xp, \Bp)$ if 
$$(\mbf{T} , \ttt ) \models \vf $$
for every $\mbf{T}$ on $(\Xp_\top , \Bp)$.
\end{definition}

\begin{notation}
Fix a $\Gamma \subseteq \mcal{L}$, let $\mbf{T}$ be a type structure on $(\Xp_\top , \Bp)$. Then we write $(\mbf{T} , \ttt) \models \Gamma$ if $(\mbf{T} , \ttt ) \models \psi$ for every $\psi \in \Gamma$.
\end{notation}

\begin{definition}[Model of $\Gamma$]
Fix a $\Gamma \subseteq \mcal{L}$. The subset $\Gamma$ has a \emph{model} in $\mfrak{T}$ if there is a type structure $\mbf{T}$ on $(\Xp_\top , \Bp)$ and a $\ttt \in \T$ such that $(\mbf{T} , \ttt) \models \Gamma$.
\end{definition}

\begin{notation}
Fix a $\Gamma \subseteq \mcal{L}$ and let $\vf \in \mcal{L}$ be arbitrary. Then we write $\Gamma \models \vf$ if 
\begin{equation*}
(\mbf{T} , \ttt ) \models \Gamma \Longrightarrow (\mbf{T} , \ttt) \models \vf
\end{equation*}
for every type structure $\mbf{T}$ on $(\Xp_\top , \Bp)$ and for every $\ttt \in \T$.
\end{notation}

%% file: Text/Belief-Completeness/strong_soundness_completeness/intro.tex
\subsection{Strong Soundness and Strong Completeness}
\label{subsec:strong_soundness_completeness}

%% file: Text/Belief-Completeness/strong_soundness_completeness/strong_soundness.tex
\subsubsection{Strong Soundness}
\label{subsubsec:strong_soundness}

Before introducing our system, we introduce the following piece of notation. Given a formula $\vf \in \Fin$, we let $\mathsf{Prim} ( \vf)$denote the conjunction of all the propositions from  $\otbar{\mathsf{X}}$ that appear in $\vf$.\footnote{Observe that usually this set is defined according to the set of primitive propositions $\mathsf{X}$ alone. However, since our conditioning propositions are defined on $\otbar{\mathsf{X}}$, we need to modify the definition accordingly.}

The following is the list of axioms schemata and inference rules of our system $H_B$. Observe that the list is not minimal.

\begin{definition}[System $H_B$]
\label{def:HB}
The \emph{system} $H_B$ is given by the constant $\top$ and by the following list of axioms schemata and inference rules (where every  $\alpha, \beta \in \bQp$). 
\begin{itemize}[leftmargin=*]
\item \emph{Axioms schemata}:
\begin{itemize}[leftmargin=*]
\item Primitive Propositions Schemata: 
\begin{enumerate}[label=A\arabic*., leftmargin=*]
\item $\varphi \rightarrow ( \psi \rightarrow \varphi ) ,$ \hfill $\forall \varphi, \psi \in \mcal{L}$;
\item $(\varphi \rightarrow ( \psi \rightarrow \rho ) ) \rightarrow ((\varphi \rightarrow \psi ) \rightarrow ( \varphi \rightarrow \rho ) ) ,$ \hfill $\forall \varphi, \psi, \rho \in \mcal{L}$;
\item $(\neg \varphi \rightarrow \neg \psi ) \rightarrow ( \psi \rightarrow \varphi ),$ \hfill $\forall \varphi, \psi \in \mcal{L}$;
\item $\displaystyle \bigwedge_{\varphi \in \Phi} ( \psi \rightarrow \varphi )  \rightarrow \psi \rightarrow  \bigwedge_{\varphi \in \Phi} \varphi  ,$  \hfill $\forall \psi \in \mcal{L} \ \forall \Psi \subseteq \mcal{L} \ ( \abs{\Psi} \leq 2^{\aleph_\gamma} )$;
\item $\displaystyle \bigwedge_{\varphi \in \Phi} \varphi \rightarrow \psi ,$  \hfill $\forall \psi \in \mcal{L} \ \forall \Psi \subseteq \mcal{L} \ ( \abs{\Psi} \leq 2^{\aleph_\gamma} )$;
\item $\displaystyle \bigwedge_{a \in A} \bigvee_{b \in A} \varphi_{a, b}  \rightarrow  \bigvee_{g \in A^A} \bigwedge_{a \in A} \varphi_{a, g(a)}  ,$ \hfill $\forall \varphi \in \mcal{L}_0, \ \abs{A} \leq \aleph_\gamma$;
\end{enumerate}
\item Conditional Probabilistic Schemata:
\begin{enumerate}[label=P\arabic*., leftmargin=*, itemsep=0.5ex]
\item $\displaystyle \p^{0}_{j , \F} (\vf)$, \hfill $ \forall \F \in \Bp \ \forall \vf \in \Fin$;
\item $\displaystyle \p^{1}_{j , \F} (\top)$, \hfill $ \forall \F \in \Bp$;
\item $\displaystyle \p^{1}_{j , \F} (\F)$, \hfill $ \forall \F \in \Bp $;
\item $\displaystyle \p^{1}_{j , \F} (\mathsf{Prim} (\vf))$, \hfill $ \forall \F \in \Bp \ \forall \vf \in \Fin ( \mathsf{Prim} (\vf) \subseteq \F )  $;
\item $\displaystyle \bigwedge_{\a < \beta} \p^{\a}_{j, \Phi} (\varphi)  \rightarrow \p^{\beta}_{j, \Phi} (\varphi) ,$ \hfill $\forall \F \in \Bp \ \forall \vf \in \Fin$;
\item $\Big( \p^{\a}_{j, \Phi} ( \varphi \wedge \psi) \wedge \p^{\beta}_{j, \Phi} ( \varphi \wedge \neg \psi ) \Big) \rightarrow \p^{\alpha + \beta}_{j, \Phi} (\varphi) ,$ \hfill $\forall \vf,  \psi \in \Fin , \ \alpha + \beta \leq 1 $;
\item $\Big( \neg \p^{\a}_{j, \Phi} ( \varphi \wedge \psi) \wedge \neg \p^{\beta}_{j, \Phi} ( \varphi \wedge \neg \psi ) \Big) \rightarrow \neg \p^{\alpha + \beta}_{j, \Phi} (\varphi) ,$ \hfill $\forall \vf,  \psi \in \Fin , \  \alpha + \beta \leq 1 $;
\item $\p^{\a}_{j, \Phi} ( \varphi)  \rightarrow \neg  \p^{\beta}_{j, \Phi} (\neg \varphi) ,$ \hfill $\forall \vf \in \Fin , \ \alpha + \beta > 1 )$;
\item $\p^{1}_{j, \Phi} ( \varphi \rightarrow \psi ) \rightarrow (\p^{\a}_{j, \Phi} (\varphi) \rightarrow \p^{\a}_{j, \Phi} ( \psi ) ) ,$ \hfill  $\forall \vf \in \Fin$;
\end{enumerate}
\item Updating Schemata:
\begin{enumerate}[label=U\arabic*., leftmargin=*, itemsep=0.5ex]
\item $\Big( \p^{\alpha}_{j, \Psi} ( \Xi) \wedge %
\p^{\beta}_{j, \F} ( \Psi) \Big) \rightarrow %
\p^{\alpha\beta}_{j, \F} ( \Xi) ,$ %
\hfill $\forall \F \subseteq \Psi \subseteq \Xi \in \Bp$;
\item $\Big( \p^{\alpha}_{j, \Psi} ( \vf) \wedge %
\p^{\beta}_{j, \F} ( \Psi) \Big) \rightarrow %
\p^{\alpha\beta}_{j, \F} ( \vf) ,$ %
\hfill $\forall \F \subseteq \Psi \in \Bp \ \forall \varphi \in \Fin$;
\item $\Big( \neg  \p^{\alpha}_{j, \Psi} ( \Xi) \wedge %
\neg \p^{\beta}_{j, \F} ( \Psi) \Big) \rightarrow %
\neg \p^{\alpha\beta}_{j, \F} ( \Xi) ,$ %
\hfill $\forall \F \subseteq \Psi \subseteq \Xi \in \Bp$;
\item $\Big( \neg  \p^{\alpha}_{j, \Psi} ( \vf) \wedge %
\neg \p^{\beta}_{j, \F} ( \Psi) \Big) \rightarrow %
\neg \p^{\alpha\beta}_{j, \F} ( \vf) ,$ %
\hfill $\forall \F \subseteq \Psi \in \Bp \ \forall \varphi \in \Fin$;
\end{enumerate}
\end{itemize}
\item \emph{Inference rules:}
\begin{itemize}[leftmargin=*]
\item (MP) Modus Ponens: From $\varphi$ and $\varphi \rightarrow \psi$ infer $\psi$;
\item (C) Conjunction: From $\Gamma \subseteq \mcal{L}$ such that $\abs{\Phi} \leq 2^{\aleph_\gamma}$ infer $\displaystyle \bigwedge_{\vf \in \Gamma} \varphi$;
\item (N) Necessitation: From $\vf \in \Fin$ infer $\p^{1}_{j , \F} (\vf)$, for every $\F \in \Bp$;
\item (C0) Continuity at $\varnothing$: From $\displaystyle \bigwedge_{n \in \bN} \vf_n \rightarrow \neg \top$ where, for every $n \in \bN$, $\vf_n \in \Fin$, infer, for every $\F \in \Bp$, 
\begin{equation*}
\bigwedge_{n \in \bN \setminus \{ 0 \}} \Bigg( \bigvee_{l \in \bN} \neg \p^{\frac{1}{k}}_{j , \F} \Bigg( \bigwedge_{n \leq l} \vf_n \Bigg) \Bigg) .
\end{equation*}
\end{itemize}
\end{itemize}
If additionally $\aleph_\gamma > \aleph_0$, then $H_B$ includes: 
\begin{itemize}[leftmargin=*]
\item \emph{Axiom schemata}:
\begin{itemize}[leftmargin=*]
\item Introspection Schema:
\begin{enumerate}[label=I\arabic*., leftmargin=*, itemsep=0.5ex]
\item $\p^{\a}_{j , \F} (\varphi) \rightarrow %
\p^{1}_{j , \F} ( \p^{\a}_{j , \F} (\varphi) )$, %
\hfill $\forall \F \in \Bp \ \forall \vf \in \Fin$;
\end{enumerate}
\end{itemize}
\item Inference Rule:
\begin{itemize}[leftmargin=*]
\item (UI) Uncountable Introspection:  For every $j \in I$, from $\displaystyle \varphi \rightarrow \Bigg( \bigvee_{n \in \bN} \varphi_n \Bigg)$, with $\varphi \in \mcal{L}_j$ and for every $n \in \bN$, $\varphi_n \in \Fin$, infer, for every $\F \in \Bp$, %
\begin{equation*}
\varphi \rightarrow \bigwedge_{k \in \bN \setminus \{ 0 \} } \Bigg( \bigvee_{ l \in \bN } \p^{1 - \frac{1}{k}}_{j , \F} \Bigg( \bigvee_{n \leq l} \varphi_n \Bigg) \Bigg) .
\end{equation*}
\end{itemize}
\end{itemize}
\end{definition}

As it is customary the \emph{set of theorems} of $H_B$ is the smallest set of formulae that contains the objects in \Mref{def:HB}. Also, a \emph{proof} of $\varphi$ from $\Gamma$ in the system $H_B$ is a sequence such that:
\begin{itemize}[leftmargin=*]
\item the length of the sequence is strictly smaller than $2^{\aleph_\gamma}$ \item $\varphi$ is the last formula of the sequence, 
\item every formula present in the sequence is either a theorem of $H_B$ or it is a formula inferred from the previous formulae via Modus Ponens of Conjunction.
\end{itemize}
Finally, we say that a subset $\Gamma \subseteq \mcal{L}$ \emph{implies syntactically} $\varphi \in \mcal{L}$, written $\Gamma \vdash \varphi$, if there is a proof of $\varphi$ from $\Gamma$.

\begin{definition}[Consistent Family of Formulae]
A set of formulae $\Gamma$ is \emph{consistent}, written $\mfrak{C} ( \Gamma)$, if there is no formula $\vf \in \mcal{L}$ such that there are proofs of $\varphi$ and $\neg \vf$ from $\Gamma$ in $H_B$.
\end{definition}

\begin{definition}[Strong Soundness]
The system $H_B$ is \emph{strongly sound} if 
\begin{equation*}
\Gamma \vdash \varphi \Longrightarrow \Gamma \models \varphi
\end{equation*}
for every $\Gamma \subseteq \mcal{L}$ and for every $\varphi \in \mcal{L}$.
\end{definition}

\begin{proposition}[Strong Soundess of $H_B$]
\label{prop:strong_soundness}
The system $H_B$ is strongly sound with respect to the class of type structures $\mfrak{T}$.
\end{proposition}

%% file: Text/Belief-Completeness/strong_soundness_completeness/strong_completeness.tex
\subsubsection{Strong Completeness}
\label{subsubsec:strong_completeness}

First we recall for self-containment the definition of strong completeness.

\begin{definition}[Strong Completeness]
The system $H_B$ is \emph{strongly complete} if 
\begin{equation*}
\Gamma \models \varphi \Longrightarrow \Gamma \vdash \varphi
\end{equation*}
for every $\Gamma \subseteq \mcal{L}$ and for every $\varphi \in \mcal{L}$.
\end{definition}

The following is the definition of what we call the \emph{canonical} (measurable) type spaces that comprise the object  we are after. 

\begin{definition}[The Canonical Type Spaces]
\label{def:canonical_type}
Fix a tuple $(\Xp_\top , \Bp)$ and let $( S, \Sigma_S, \mcal{B} )$ be the conditional measurable space induced from $(\Xp_\top , \Bp)$. Then:
\begin{itemize}[leftmargin=*]
\item for every $i \in I_0$, define
\begin{equation*}
\T^{*}_i := \Set { %
\bigwedge_{\varphi \in \Theta_{j}} \varphi \wedge %
\bigwedge_{\psi \in \Fin_{j} \setminus \Theta_{j}} \neg \psi | %
\Theta_{j} \subseteq \Fin_{j} : %
\mfrak{C} \Round { \Theta_{j} \cup \Set { \neg \psi | \psi \in \Fin_{j} \setminus \Theta_{j} } } } ;
\end{equation*}
\item define $\T^* := \prod_{i \in I_0} \T^{*}_i$.
\end{itemize}
For every $i \in I_0$ and $\psi_i \in \Fin_{i}$ define
\begin{equation*}
[\psi_i]_i := \Set { \ttt_i \in \T^{*}_i | \vdash \ttt_i \rightarrow \psi_i }
\end{equation*}
Then,
\begin{itemize}[leftmargin=*]
\item the $\sigma$-algebra on $T^{*}_i$ is defined as 
\begin{equation*}
\Sigma^{*}_i := \sigma \Round { \Set { [\psi_i]_i | \psi_i \in \Fin_{j} } }
\end{equation*}
for every $i \in I_0$;
\item the $\sigma$-algebra on $T^*$ denoted by $\Sigma^*$, is the product $\sigma$-algebra of the $\sigma$-algebras $\Sigma^{*}_i$, with $i \in I_0$.
\end{itemize}
\end{definition}

\begin{notation}
We let $C_{i}^{*} := C^{*}_i \times \prod_{y \in I_0 \setminus \{ i \} } \Omega^{*}_y$ with $C_i \in \Sigma^{*}_i$, for every $i \in I_0$. Also, we let $[\vf_i]^* := ([\vf_i ]_i )^*$. 
\end{notation}

The definitions that follow introduce the belief functions and the valuation function. Much in the same spirit of \Sref{sec:terminality} we have to prove that the belief functions are actually measurable.

\begin{definition}[The Belief Functions]
\label{def:canonical_belief}
The \emph{canonical belief function} $\m^{*}_j$, for every $j \in I$, is defined as a profile $\m^{*}_j := ( \m^{*}_{j, B} )_{B \in \mcal{B}}$, where $ \m^{*}_{j, B}$, for every $B \in \mcal{B}$, is defined as 
\begin{equation*}
\m^{*}_{j, B} ( \ttt_j ) ( [\vf]^* ) := \sup \Set { \alpha \in \bQp | \vdash \ttt_j \rightarrow \p^{\alpha}_{j , \F} ( \vf ) }
\end{equation*}
for every $\F \in \Bp$ which corresponds to $B \in \mcal{B}$ and $\vf \in \Fin$.
\end{definition}

\begin{definition}[The Valuation Function]
\label{def:canonical_valuation}
The \emph{canonical valuation function} $\val^{*}$ is defined as 
\begin{equation*}
\val^* ( s , \mathsf{p} ):=
\begin{cases}
1, & \text{ if } s \in [\mathsf{p}],\\
0, & \text{ if } s \notin [\mathsf{p}],
\end{cases}
\end{equation*}
for every $s \in S$, and for every $\mathsf{p} \in \Xp$, and
\begin{equation*}
\val^* ( s, \top ) =1
\end{equation*}
always.
\end{definition}

We are finally in position to define the canonical type structure $\mbf{T}^*$ on $(\Xp_\top , \Bp)$, which is comprised of all the objects previously introduced in this section.

\begin{definition}[The Canonical Type Structure $\mbf{T}^*$]
\label{def:canonical_terminal}
The \emph{canonical type structure} $\mbf{T}^*$ on $(\Xp_\top, \Bp)$ is the tuple
\begin{equation*}
\mbf{T}^* := \la I_0 , ( \Xp_\top, \Bp ), (S, \Sigma_S , \mcal{B}) , (\T^{*}_i )_{i \in I_0 } , (\m^{*}_j )_{j \in I} , \val^{*} \ra
\end{equation*}
where 
the type spaces $\T^{*}_i$, for every $i \in I_0$ are defined as in \Mref{def:canonical_type}, the belief functions $\m^{*}_j$, for every $j \in I$, are defined as in \Mref{def:canonical_belief}, and the valuation function $\val^{*}$ is defined as in \Mref{def:canonical_valuation}.
\end{definition}

The following result is crucially established via the canonical type structure as defined in \Mref{def:canonical_terminal}.

\begin{proposition}[Strong Completeness of $H_B$]
\label{prop:strong_completeness}
The system $H_B$ is strongly complete with respect to the class of type structures $\mfrak{T}$.
\end{proposition}

%% file: Text/Belief-Completeness/belief-completeness.tex
\subsection{Belief-Completeness}
\label{subsec:belief-completeness_non-redundancy}

The canonical type structure $\mbf{T}^*$ constructed in the previous section is nothing more than the terminal type structure $\mscr{T}^* := \la I_0 , ( S, \Sigma_S, \mcal{B}) , (T^{*}_i )_{i \in I_0 } , (m^{*}_j )_{j \in I} \ra$ previously constructed in \Sref{subsec:terminal_type_infinite}. This is indeed established next.

\begin{proposition}[Terminality of $\mbf{T}^*$]
\label{prop:terminality_valuation_type}
The type structure $\mbf{T}^* := \la I_0 , ( \Xp_\top, \Bp ), (S, \Sigma_S , \mcal{B}) , (\T^{*}_i )_{i \in I_0 } , (\m^{*}_j )_{j \in I} , \val^{*} \ra$ is terminal.
\end{proposition}

\begin{corollary}
\label{cor:sameness}
From \Mref{rem:uniqueness_terminality}, the terminal type structure $\mscr{T}^*$ constructed in \Sref{subsec:terminal_type_infinite} and the terminal type structure $\mbf{T}^*$ constructed in \Mref{def:canonical_terminal} are one and the same object.
\end{corollary}

Before stating the main result of this section, one last piece of notation, which let us express the belief function in a version often encountered in the literature.\footnote{See for example \cite{Battigalli_Siniscalchi_2002} and \cite{Brandenburger_et_al_2008}, or \cite{Dekel_Siniscalchi_2015}.}

\begin{notation}
For every $j \in I$, let $\beta_j := \marg_{\T_{-j}} \circ \ m_j$, that is, 
\begin{equation*}
\beta_j :=   (\beta_{j, B})_{B \in \mcal{B}}) : T_j \to \Delta^{\mcal{B}} (T_{-j}).
\end{equation*}
The function $\beta^{*}_j$ is defined accordingly with the obvious modifications.
\end{notation}

\begin{theorem}[Belief-Completeness of $\mbf{T}^*$]
\label{th:belief-completeness}
In the $H_B$ system, let $j \in I$ be arbitrary with $\mu^{*}_j \in \Delta^{\mcal{B}} (\T^{*}_{-j})$. Thus, there is one and only one $\ttt_j \in \T^{*}_j$ such that $\beta^{*}_j (\ttt_j) = \mu^{*}_j$. Also, for every $j \in I$, $\beta^{*}_j$ is a measurable isomorphism.
\end{theorem}

Hence, we are now in position to state the final result of the paper as a theorem, even if it is simply a corollary of  \Mref{th:universal_hierarchies}, \Mref{th:non-redundancy}, and \Mref{th:belief-completeness}. In doing so, we are allowed from \Mref{cor:sameness} to move back to the notation employed in \Sref{sec:terminality} and \Sref{sec:non-redundancy}.

\begin{theorem}[Universality of $\mscr{T}^*$]
\label{th:universality}
The type structure $\mscr{T}^* := \la I_0 , ( S, \Sigma_S, \mcal{B}) , (T^{*}_i )_{i \in I_0 } , (m^{*}_j )_{j \in I} \ra$ is the universal type structure up to measurable isomorphism.
\end{theorem}

%% file: Background/bibliography.tex

\bibliography{./Background/Biblio_Infinitary_TARK.bib}

\addcontentsline{toc}{section}{References}

\bibliographystyle{./Background/eptcs.bst}